\documentclass[12pt,a4]{article}

\usepackage[pdftex]{hyperref}
\usepackage{amsthm,amsmath,latexsym,amssymb,amsfonts}
\usepackage{graphicx,lscape,fancyhdr,array,stmaryrd,euscript,wrapfig}

\begin{document}
\begin{titlepage}

\begin{flushright}
\vspace{1mm}
\end{flushright}

\vspace{1cm}
\begin{center}
{\bf \Large Higher Spin Fields and Symplectic Geometry}
\vspace{2cm}

\textsc{ Dmitry
Ponomarev\footnote{dmitri.ponomarev@umons.ac.be}}

\vspace{2cm}

{\em Service de M\'ecanique et Gravitation, Universit\'e de Mons -- UMONS\\
20 Place du Parc, 7000 Mons (Belgium)}

\end{center}

\vspace{0.5cm}
\begin{abstract}

We argue that higher spin fields originate from 
Hamiltonian mechanics and play a role of gauge fields 
ensuring covariance of geometric observables such as 
length and volume  with respect 
to canonical transformations in the same way as 
a metric tensor in Riemannian geometry
 ensures covariance with respect to 
diffeomorphisms. We consider a  
reparametrization invariant action of a point particle
in  Hamiltonian form.
Reparametrization invariance is achieved in the standard 
way by coupling to the auxiliary world-line metric.
 Identifying Hamiltonian function
with a generating function for higher spin fields
this action can be viewed as an action for the point 
particle in a higher spin background, while canonical
transformations act as higher spin symmetries.
We define the gauge invariant length as 
a proper time of a particle moving along the geodesic.
Following the usual geometrical interpretation we introduce
the volume form and the scalar curvature for a combined lower spin sector.
As for the general case, we show that notions of local volume and 
scalar curvature are not compatible with symplectic transformations.
We propose symplectically invariant
 counterparts for the total volume of the
space and Einstein-Hilbert action.

\end{abstract}

\end{titlepage}

\numberwithin{equation}{section}

\section{Introduction}
\label{intro}

Besides the underlying role played in general relativity, 
(pseudo)-Riemannian geometry exists independently 
 as an inherent part of numerous branches 
of modern physics and geometry. Taking metric as an exhausting
characteristic
of curved space
it allows to treat notions of length, 
angle, geodesic, parallel transport, volume, curvature, etc. 
In turn, general relativity claims that the metric is dynamical,
it is governed by non-linear equations of motion and can be sourced 
by other fields.
From this perspective it is to be expected that 
higher spin theory\footnote{For recent reviews on higher spin gauge 
theories see e.g. \cite{Sorokin:2004ie,Bekaert:2005vh,Bekaert:2010hw,Sagnotti:2011qp}}
, being an extension of usual gravity,
should not only be an interacting theory of higher spin fields,
but it should also  preserve or deform properly underling geometrical 
concepts listed above.

 Despite the success of Vasiliev's theory
  \cite{Vasiliev:1990en,Vasiliev:2003ev} as a theory of 
interacting higher spin fields consistent at all orders in coupling constant,
the geometrical interpretation of higher spin fields remains vague.
Neither length nor curvature are gauge invariant with respect 
to higher spin transformations. Absence
 of gauge invariant geometric observables
 leads to significant difficulties in interpretation
of solutions to higher-spin equations
\cite{Didenko:2006zd,Didenko:2009td,Gutperle:2011kf,Iazeolla:2011cb,Castro:2011iw,Iazeolla:2012nf,Ammon:2012wc}.
For instance, the position of the horizon of the higher-spin black hole   cannot 
be defined because the usual interval 
changes under higher-spin gauge transformations.

The other role of the gravity field is  that it insures 
covariance with respect to diffeomorphisms via replacing usual derivatives
with covariant ones, etc. In contrast to internal symmetries, 
 diffeomorphisms are transformations of the space-time, which makes
 difficult to unify gravity with other interactions. This conceptual 
 problem can be resolved via the frame-like approach
 where  gravity enters in Cartan's form. It is believed to be inevitable
 when one needs to deal with fermions. 
 In the frame-like approach one introduces locally a basis where the
 metric is Minkowskian.
 Then  gravity becomes a gauge theory of local Poincare or $SO(d-1,2)$ symmetry acting 
 in a fiber space, while diffeomorphisms covariance is ensured by 
 usage of differential forms.
In this way the frame-like approach erases any difference between 
gravity   
and other fields treating all of them as gauge fields for local internal symmetries.
 In particular, in Vasiliev's theory diffeomorphisms
are no longer associated with gravity, being shared in the same extent by all the higher spin
fields.

In this paper we chose the opposite option.
 Namely, we argue
that 
the higher spin extension of diffeomorphisms should be 
viewed as a group of transformations of the space, which obviously, should be 
enlarged with new coordinates, since diffeomorphisms already exhaust all 
the transformations of the ordinary space-time. The proper extension
is an extension of coordinates with momenta, which leads to the phase space, while 
diffeomorphisms become enlarged to canonical transformations also called 
symplectic diffeomorphisms or symplectomorphisms. One 
can similarly consider quantum analogue of canonical transformations, 
forming Heisenberg group, but we restrict ourselves to a 
more simple classical case. Let us mention that canonical transformations in one way or another appeared
in the context of higher spin theories
 \cite{Berends:1984rq,Vasiliev:1990en,Segal:2000ke,Segal:2002gd,Vasiliev:2003ev,Vasiliev:2005zu,Engquist:2005yt,Grigoriev:2006tt,Bekaert:2009ud,Bekaert:2010ky,NeveuLPI,Grigoriev:2012xg}.
The very fact that higher spin transformations acting on higher spin fields can be thought
of as transformation of Hamiltonian function induced by symplectomorphisms
gives a strong evidence that higher spin symmetries should be treated exactly 
from this point of view. 

The setting where one deals with the phase space in the 
 context of higher spin 
theory and identifies Hamiltonian function and canonical transformations 
with the generating functions for higher spin fields and
symmetries  respectively has been used in \cite{Bars:1998gv,Segal:2000ke,Segal:2002gd,Grigoriev:2006tt,Bekaert:2009ud,Bekaert:2010ky}. In particular, in 
\cite{Segal:2002gd}
fully non-linear conformal higher spin theory has been built.

There are several well-known facts
from Hamiltonian and quantum mechanics, which imply
that coordinates and momenta should be treated on the same footing:
\begin{itemize}
\item the full set of the initial data is given by initial coordinates and 
momenta;
\item coordinates and momenta appear in canonical commutation relations in a
symmetric way;
\item phase-space admits a natural volume form;
as a result, the path integral measure for the Lagrangian formulation 
of a point particle
can be derived only from the Hamiltonian one. 
\end{itemize}
Thus, both the phase space with canonical transformations acting in it and Riemannian
geometry are the inherent parts of physics. Regardless of the connection with higher
spin fields, it seems natural to try to unify them by making a generalization of Riemannian geometry,
that is
covariant under canonical transformations.

The aim of this paper is to construct some geometrical observables of 
higher spin geometry generalizing respective notions of (pseudo)-Riemannian 
geometry. 
In Section \ref{ppaction} we discuss the action of a point particle interacting 
with a background higher spin field introduced in 
\cite{Segal:2000ke}\footnote{The problem of construction of the point particle action 
invariant with respect to higher spin transformations has been addressed 
 and solved in the first non-trivial  order in higher spin
fields in \cite{deWit:1979pe}.}. It is the most general action, that depends 
on coordinates and velocities only, while the reparametrization invariance is achieved 
in the standard way by coupling to an auxiliary world-line metric. Written in 
a Hamiltonian form it is manifestly invariant with respect to canonical transformations,
which are identified as  higher-spin symmetries. In the same time Hamiltonian should be 
thought of as  the generating function for all higher-spin fields,
which is inferred in Section \ref{hs}. 
 The proper time measured by 
the inner world-line metric provides a manifestly invariant 
definition of length.
This is discussed and exemplified in Section \ref{length}.
Let us note that the reparametrization invariance of the point
particle action implies an extra symmetry acting on Hamiltonian.
It was called hyper-Weyl symmetry
in \cite{Segal:2000ke}. Presence of this symmetry in the higher
spin action eventually results in a conformal higher spin theory not a
massless one.
 The length introduces a distinguished 
parametrization and thereby breaks this symmetry.
All the notions that we are going to construct next 
essentially rely on the length.
In this sense,
 the length is the additional structure that allows to construct a wider set
of natural 
symplectic invariants not necessary being invariant with respect to
hyper-Weyl symmetry.

In  Section \ref{volume} we introduce a volume form 
 so as 
it agrees with the length, namely, we require that the
volume of an infinitesimal geodesic ball depends on its
radius in the same way as in the flat case. 
This enables us to introduce a scalar curvature in Section \ref{curvature}
in a conventional way extracting it from the volume defect of a geodesic ball 
in subleading orders in radius.
Since generically canonical transformations mix coordinates with momenta, 
the volume of the domain in $x$ space is not well-defined quantity.
So, there are no reasons to expect that the  curvature defined is invariant 
and can be used to construct invariant action for higher spin fields.
Nevertheless, this curvature is manifestly scalar with respect to 
diffeomorphisms and electromagnetic gauge transformations
and thereby allows to build an invariant action for spins 0, 1 and 2, which directly generalizes 
Einstein-Hilbert action to the sector
of lower spin fields.
It is to be expected that this action
should appear as a lower spin truncation of the full one.
The lower spin action is computed in Section \ref{curvature}.

In  Section \ref{invariants} we discuss symplectic invariants, that
are given in terms of lengths of closed geodesics. After considering
a simple example we propose two such invariants as candidates
for higher spin generalizations of the total volume and Einstein-Hilbert action.

\section{Point particle action}
\label{ppaction}

Let us consider the most general reparametrization invariant action of a point particle
that depends only on coordinates and velocities
\begin{equation}
\label{lagact}
S=\int{L(x,u)}edt,
\end{equation}
where $e(t)$ is a world-line metric and
\begin{equation*}
u^i=\frac{dx^i}{edt}=\frac{\dot x^i}{e}.
\end{equation*}
 Under the reparametrization of the world-line
$t\rightarrow t'$ the world-line metric transforms so as the interval
\begin{equation}
\label{length1}
ds=e(t'(t))dt'=e(t)dt
\end{equation}
remains constant,
 which ensures 
reparametrization invariance. The same action can be transformed to the Hamiltonian 
form 
\begin{equation}
\label{int1}
S=\int{(p_idx^i-\tilde{H}(p,x)dt)},
\end{equation}
where $\tilde{H}$ is a Legendre transform of the Lagrangian
\begin{equation}
\label{int2}
\tilde{L}(x,\dot x)=L(x,\frac{\dot x}{e})e,
\end{equation}
that is
\begin{equation}
\label{int3}
\tilde{H}(p,x)=p_i\dot x^i-\tilde{L}(x,\dot x(p,x))
\end{equation}
and $\dot x$ is expressed in terms of $x$ and $p$ via
\begin{equation}
\label{int4}
p_i=\frac{\partial \tilde{L}}{\partial \dot x^i}(x,\dot x).
\end{equation}

It is easy to check, that the action (\ref{int1}) admits an equivalent form
\begin{equation}
\label{hamact}
S=\int{(p_idx^i-H(p,x)edt)},
\end{equation}
where $H$ is a Legendre transform of $L(x,u)$, that is
\begin{equation*}
H(p,x)=p_iu^i(p,x)-L(x,u(p,x)), 
\end{equation*}
and $u(p,x)$ solves
\begin{equation*}
 p_i=\frac{\partial L}{\partial u^i}(x,u),
\end{equation*}
which is equivalent to (\ref{int4}).

The action (\ref{hamact}) up to boundary terms is invariant with respect to canonical transformations,
which infinitesimally are
\begin{equation}
\label{canontr1}
\delta x^i =[\varepsilon(p,x), x^i], \quad \delta p_i =[\varepsilon(p,x),p_i],
\end{equation}
\begin{equation}
\label{canontr2}
\delta H(p,x)=-[\varepsilon(p,x),H(p,x)]
\end{equation}
with 
\begin{equation}
\label{poi}
[f,g]=\frac{\partial f(p,x)}{\partial p_i}\frac{\partial g(p,x)}{\partial x^i}-
\frac{\partial f(p,x)}{\partial x^i}\frac{\partial g(p,x)}{\partial p_i}.
\end{equation}

Let us note that $H$ transforms as a scalar under the action of canonical 
transformations in the sense that 
\begin{equation*}
\delta H(p,x)=H'(p,x)-H(p,x) \quad {\text{where}} \quad 
H'(p+\delta p,x+\delta x)=H(p,x).
\end{equation*}
This is not to be confused with the 
increment $\delta f(p,x)$ of a fixed function $f(p,x)$ produced
by the change of its arguments $(p,x)\rightarrow(p+\delta p,x+\delta x)$
\begin{equation*}
\delta f(p,x)=f(p+[\varepsilon, p],x +[\varepsilon, x])-f(p,x)= [\varepsilon(p,x),f(p,x)],
\end{equation*}
which gives the opposite sign compared to (\ref{canontr2}). 

 Transformations (\ref{canontr1}), (\ref{canontr2}) 
can be translated into lagrangian representation, thus giving transformations acting
on coordinates, velocities and Lagrangian function. 

 The action (\ref{hamact}) also enjoys hyper-Weyl symmetry
\begin{equation}
\label{hyper}
H'=A(x,p)H, \quad e'=A^{-1}(x,p)e.
\end{equation}

\section{Higher spin fields}
\label{hs}

Our aim is to show that one can regard $H$ and $L$ as generating functions for higher spin fields, while
(\ref{lagact}), (\ref{hamact}) acquire meaning of the action of a point particle in 
the higher spin background. More precisely, we identify the particular values of $H^{(0)}$ and $L^{(0)}$ 
that should be associated with the Minkowski space and show that at the linear order in 
fluctuations $H^{(1)}$ and $L^{(1)}$
\begin{itemize}
\item coefficients $h^{a(s)}(x)$ and $l_{a(s)}(x)$ \footnote{Here notation
 $a(s)$ means
that the tensor is a rank-$s$ symmetric one. We will also use notation of the form
 $\partial^a\varepsilon^{a(s-1)}$
to indicate that $s-1$ symmetric indices of $\varepsilon$ are symmetrized with the index
carried by the derivative with the strength one. For a partial derivative $\partial/\partial x^i$
we sometimes use a shortcut notation $\partial_i$. Derivatives with respect to other 
variables are written 
explicitly.}
of  Taylor expansion
\begin{equation}
\label{powers}
H^{(1)}(p,x)=\sum_{s=0}^{\infty}{\frac{1}{s!}h^{a(s)}(x)\overbrace{p_a\dots p_a}^{s}}, 
\qquad
L^{(1)}(x,u)=\sum_{s=0}^{\infty}{\frac{1}{s!}l_{a(s)}(x)\overbrace{u^a\dots u^a}^{s}}
\end{equation}
behave as rank-$s$ contravariant and covariant tensors  under diffeomorphisms respectively.
Diffeomorphisms form the subalgebra of canonical transformations (\ref{canontr2}) generated by 
$\varepsilon$, that are linear in momenta;
\item Canonical transformations (\ref{canontr2}) linearized near the background Minkowski space acting on $h^{a(s)}$ and $l_{a(s)}$
reproduce standard linear higher spin gauge transformations
 $\delta \phi^{a(s)}=\partial^a\varepsilon^{a(s-1)}$
as in Fronsdal's theory \cite{Fronsdal:1978rb},
 but with unconstrained gauge parameters and  
fields\footnote{In Fronsdal's approach \cite{Fronsdal:1978rb}
fields are supposed to be doubletracaless, while gauge parameters are traceless.
These constraints appear naturally in the frame-like approach to higher spin fields
\cite{Vasiliev:1980as,Bekaert:2005vh}. The other point of view is to let fields and gauge 
parameters be unconstrained \cite{Pashnev:1998ti,Buchbinder:2001bs,Francia:2002aa,Bekaert:2002dt,deMedeiros:2002ge,Francia:2002pt,Bekaert:2003az}. 
For a review on different approaches see \cite{Sorokin:2004ie}.}.
\end{itemize}
This allows to interpret $h^{a(s)}$ and $l_{a(s)}$ as off-shell spin-$s$ field.  
Both ways to identify the higher spin field are legitimate, moreover, as we will
illustrate,  for small fluctuations around Minkowski space
 $h^{a(s)}$ and $l_{a(s)}$ are related by  Legendre transform, which
up to a sign factor coincides with the usual convention for raising and lowering
indices via background Minkowski metric. However, for general field
configuration relation between $h^{a(s)}$ and $l_{a(s)}$ is far not as 
straightforward because it implies solving velocities in terms of momenta or vice versa.
One should not be bothered by this fact because notion of spin is defined
only in vicinity of Minkowski space. For general field configuration the only
natural way to deal with higher spin fields is to consider all them
 within one generating
function $H$ or $L$.

Before showing this explicitly let us make the following comment. Since all the higher 
spin fields are unified by the symmetry group, the point particle can have only one 
charge with respect to this group. It appears as an overall factor in front of 
the point particle action and can be omitted for  brevity. In nature, however,
particle charges with respect to fields of different spin,
such as mass and electric charge, can vary independently. 
Without breaking the symmetry of the theory this can happen if particles 
appear as solitonic solutions and thereby charges show up only as integration 
constants of particular solutions.

In  Minkowski space 
the action of a point particle  is 
\begin{equation}
\label{Mink}
S=\int{\left(\frac{1}{2}\eta_{mn}\frac{d x^m}{ed t}\frac{d x^m}{ed t}-
\frac{1}{2}m^2\right) edt},
\end{equation}
where we use the mostly plus convention for Minkowski metric $\eta$.
Eliminating  $e$ through its equations of motion 
it can be put into more recognizable form
\begin{equation*}
S=-m\int{\sqrt{-\eta_{mn}dx^mdx^n}}.
\end{equation*} 
As it was discussed, without loss of generality we can set $m$ 
to unity.
Comparing (\ref{Mink}) with (\ref{lagact}) we can deduce that for Minkowski 
space the background values of gravitational and scalar fields should be nonzero
\begin{equation}
\label{Mink1}
L^{(0)}(x,u)=\frac{1}{2}g^{(0)}_{mn}u^mu^n+\phi^{(0)} 
\quad \Leftrightarrow
\quad
H(p,x)=\frac{1}{2}g^{(0)mn}p_mp_n-\phi^{(0)},
\end{equation} 
\begin{equation}
\label{bgs02}
g^{(0)}_{mn}=\eta_{mn}, \quad \phi^{(0)}=-\frac{1}{2}.
\end{equation}
If we let $\phi$ be non-constant in terms of the action (\ref{Mink}) it will be viewed 
as a particle with coordinate-dependent mass. It offers interesting possibilities
to describe dark  matter in a way similar to \cite{Milgrom:1983ca}.
 It would be interesting to treat this issue
elsewhere.

Let us consider small fluctuations around (\ref{Mink1}) such that
\begin{equation}
\label{fluct1}
L=L^{(0)}(x,u)+L^{(1)}(x,u).
\end{equation}
Then 
\begin{equation}
\label{fluct2}
p_n=\eta_{mn}u^m+\frac{\partial L^{(1)}}{\partial u^n}(x,u)
\quad \Leftrightarrow \quad u^m=\eta^{ma}p_a-\frac{\partial L^{(1)}}{\partial 
u^n}(x,p_i\eta^{ij})+o(L^{(1)}),
\end{equation}
so
\begin{equation}
\label{fluct3}
H=H^{(0)}-L^{(1)}(x,p_i\eta^{ij})+o(L^{(1)}).
\end{equation}
Introducing a small fluctuation of Hamiltonian field as
\begin{equation*}
H=H^{(0)}(p,x)+H^{(1)}(p,x)
\end{equation*}
we see that in the first order of vanishing Legendre transform maps
\begin{equation}
\label{legandre}
L^{(1)}(x, u)\quad \Leftrightarrow \quad H^{(1)}(p,x)=-L^{(1)}(x,p_i\eta^{ij}).
\end{equation}

In terms of the coefficients in the power series expansion (\ref{powers}) it implies
\begin{equation}
\label{legandre1}
l^{(1)}_{a(s)}(x) \quad \Leftrightarrow \quad h^{(1)a(s)}(x)=
-l^{(1)}_{b(s)}(x)\overbrace{\eta^{ab}\dots \eta^{ab}}^s.
\end{equation}
One can readily compute from (\ref{canontr2}), that under diffeomorphisms,
which are generated by
parameters $\varepsilon$ linear in momenta $\varepsilon(p,x)=\varepsilon(x)^ip_i$ 
\begin{equation}
\label{diffeo}
\delta h^{a(s)}=-\varepsilon^b\partial_bh^{a(s)}+s\partial_{m}\varepsilon^ah^{ma(s-1)}.
\end{equation}
Eq. (\ref{diffeo}) implies that $h^{a(s)}$
manifests itself as a contravariant rank-$s$ tensor
\begin{equation*}
\delta h^{a(s)}=-{\cal L}_{\varepsilon}h^{a(s)},
\end{equation*}
where ${\cal L}_{\varepsilon}h$ is a Lie derivative of a field $h$ along a vector field $\varepsilon$. In turn from (\ref{legandre1}) 
it follows that at a given level of perturbative expansion $l_{a(s)}$ 
is a rank-$s$ covariant tensor.
Moreover, decomposing (\ref{canontr2}) into perturbative
 expansion it comes out that
 \begin{equation}
 \label{hslingtr}
 \delta H^{(1)}=-[\varepsilon,H^{(0)}]-[\varepsilon,H^{(1)}]=
 {\partial_m \varepsilon}\eta^{mn}p_n+\dots
 \end{equation}
 In terms of the expansion coefficients (\ref{powers})
we reproduce the unconstrained Fronsdal-like  higher spin transformation 
\begin{equation*}
\delta h^{a(s)}=\eta^{ab}  {\partial_b}\varepsilon^{a(s-1)}.
\end{equation*}
This implies that $h^{a(s)}$ propagates correct degrees of freedom to describe the
off-shell massless spin-$s$ field.
From (\ref{legandre1}) the same holds true for $l_{a(s)}$.

So, there are two natural ways to identify higher spin fields inside the point 
particle action. The first way is to define higher spin  fields as coefficients $l_{a(s)}$ of the 
Lagrangian expansion in powers of velocities. Historically lower spin fields were introduced
exactly this way. 
The other way is to identify higher spin fields as  coefficients $h^{a(s)}$ 
of  Hamiltonian expansion in powers of momenta. The advantage of this convention 
as well as Hamiltonian formulation in general is that symmetry with respect to 
canonical transformation is manifest and has simpler form. In particular,
(\ref{diffeo}) remains true beyond the perturbative expansion around 
 Minkowski background, while 
$l^{a(s)}$ generically do not transform as tensors. As follows from (\ref{legandre1}) 
at the linear level these two definitions differ by sign in the sense that if one 
uses the standard convention for raising and lowering indices 
then one finds $h^{a(s)}=-l^{a(s)}$.
Note that the standard convention implies a
 distinguished role of the metric. In our 
setting, where all the higher spin fields are treated  on the same footing,
 raising and lowering indices via Legendre transform seems to be more motivated.

\section{Length}
\label{length}

Before defining a length of a vector in higher spin background let us 
make few comments on equations of motion of a point particle. One can always 
eliminate world-line metric by its equation of motion, however it is more
convenient to keep it and treat  as  Lagrange multiplier enforcing constraints
in the $(p,x)$ phase space or, equivalently in the $(x,u)$ space of Lagrangian approach.
 
In the Hamiltonian 
approach $e$ plays role of  Lagrange multiplier for a constraint 
\begin{equation}
\label{constr1}
H(p,x)=0.
\end{equation}
Let us denote the subset of the phase space, that satisfies (\ref{constr1})
as ${\cal M}_H$
\begin{equation}
\label{mh}
{\cal M}_H=\{(p,x):\quad H(p,x)=0\}.
\end{equation}
Variation of (\ref{hamact}) with respect to $p$ and $x$ gives usual Hamiltonian equations
\begin{equation}
\label{eqham1}
\frac{dx^i}{edt}=\frac{dx^i}{ds}=\frac{\partial H}{\partial p_i}= [H,x^i], \qquad
\frac{dp_i}{edt}=\frac{dp_i}{ds}=-\frac{\partial H}{\partial x^i}= [H,p_i]
\end{equation}
with respect to time, measured by the interval $ds=edt$. 

In turn in Lagrangian approach (\ref{lagact})
variation with respect to $e$ enforces the constraint
\begin{equation}
\label{constr2}
L(x,u)=u^i\frac{\partial L(x,u)}{\partial u^i},
\end{equation}
while the variation with respect to $x$ gives ordinary Euler-Lagrange equations
\begin{equation}
\label{eqlag1}
\frac{\partial L}{\partial x^i}=\frac{d}{ds}\Big(\frac{\partial L}{\partial u^i}\Big).
\end{equation}
The subset of $(x,u)$  satisfying (\ref{constr2}) will be denoted as ${\cal M}_L$
\begin{equation}
\label{ml}
{\cal M}_L=\{(x,u):\quad L(x,u)=u^i\frac{\partial L(x,u)}{\partial u^i}\}.
\end{equation}

In both cases (\ref{eqham1}), (\ref{eqlag1}) proper time $s$ serves as a natural non-negative parameter
of a geodesic.
So one can define the length $\Delta s$ of a geodesic $\gamma(s_i,s_f)$ with $s \in (s_i,s_f)$ 
as an interval $\Delta s=s_f-s_i$
\begin{equation}
\label{length2}
\gamma(s_f,s_i) \quad \mapsto \quad \Delta s=s_f-s_i.
\end{equation}
 Since 
the form of Hamiltonian equations (\ref{eqham1}) is preserved under canonical transformations,
the length is gauge invariant quantity. As it was discussed above, $\Delta s$ 
is reparametrization 
invariant (\ref{length1}) as well.

In a standard way this induces the length into the tangent space. Indeed, the 
geodesic passing through the point $x(0)$ with the velocity $u(0)$ in the linear approximation is
described by the equation
\begin{equation}
\label{line}
dx^i=x^i(ds)-x^i(0)=u^i(0)ds+{\cal O}(ds^2).
\end{equation}
 This allows one to assign a length $|dx|=ds$ to a 
displacement vector $dx$ at a point $x$
\begin{equation}
\label{length3}
dx^i \quad \mapsto \quad |dx^i|=ds, \quad \text{where} \quad dx^i=dsu^i \quad \text{and} 
\quad (x^i,u^i)\in{\cal M}_L.
\end{equation}
In general, it is not always possible to present $dx$ in the form $dsu$ with $u \in {\cal M}_L$
and $ds>0$.
As a result, length is defined only for time-like vectors, that is for  those, that can 
serve as a tangent vectors to  geodesics. For other vectors the 
length can be defined through
analytical continuation.  
Finally, the length is a homogeneous function of degree one
\begin{equation*}
|\alpha dx|=\alpha |dx| \quad \text{for} \quad \alpha\ge 0.
\end{equation*}
Since the length defined by (\ref{length3}) is nothing but interval, in practice one can 
solve (\ref{constr2}) for $e$ and 
compute the length  as
\begin{equation}
\label{length4}
|dx|=ds(dx)=e(x,\frac{dx}{dt})dt=e(x,dx).
\end{equation}

Let us show that the given  definitions   reproduce  known definitions
 in a lower spin case. The lower spin Lagrangian is 
 \begin{equation}
 \label{lowslag}
 S=\int{\left(\frac{1}{2}g_{mn}(x)\frac{\dot x^m}{e}\frac{\dot x^n}{e}+
 A_m(x)\frac{\dot x^m}{e}+\phi(x)\right)edt},
 \end{equation}
where $g_{mn}$, $A_m$ and $\phi$ are gravity, electromagnetic and scalar 
fields
respectively. Solving for $e$ from its equations of motion
\begin{equation}
\label{els}
e=\sqrt{\frac{-g_{mn}\dot x^m\dot x^n}{-2\phi}}
\end{equation}
 and plugging it back 
to the action one finds
 \begin{equation}
 \label{lowslag1}
 S=\int{\left(-\sqrt{-g_{mn}\dot x^m\dot x^n}{\sqrt{-2\phi}}+A_m\dot x^m\right)dt}.
 \end{equation}

The Hamiltonian form of the action (\ref{lowslag}) is
\begin{equation}
\label{lowsham}
S=\int{\left(p_i\dot x^i-e[\frac{1}{2}h^{mn}(x)p_mp_n+h^m(x)p_m+h(x)]\right)dt},
\end{equation}
where
\begin{equation}
\label{lslt}
g_{mn}=(h^{-1})_{mn}, \quad A_m=-h^n(h^{-1})_{mn}, \quad \phi=-h+\frac{1}{2}(h^{-1})_{mn}h^mh^n.
\end{equation}
The lower spin sector of canonical transformations (\ref{canontr2}) is
\begin{equation}
\label{lsgtr1}
\varepsilon(p,x)=\varepsilon^n(x)p_n+\varepsilon(x).
\end{equation}
 It consists of diffeomorphisms
$\varepsilon^n(x)$
and $U(1)$ gauge transformations $\varepsilon(x)$, which act as follows
\begin{align}
\label{s2}
\delta h^{kl}&=-\varepsilon^m\partial_mh^{kl}+\partial_n\varepsilon^kh^{nl}+
\partial_n\varepsilon^lh^{nk}=-{\cal L}_{\varepsilon}h^{kl},
\\
\label{s1}
\delta h^k&=-\varepsilon^m\partial_m h^k+\partial_m\varepsilon^k h^m+h^{mk}\partial_m\varepsilon=
-{\cal L}_{\varepsilon}h^k+h^{mk}\partial_m\varepsilon, 
\\
\label{s0}
\delta h &= -\varepsilon^m\partial_m h+h^m\partial_m\varepsilon=
-{\cal L}_{\varepsilon}h+h^m\partial_m\varepsilon
\end{align}
or, equivalently,
\begin{align}
\label{s22}
\delta g_{kl}&=-\varepsilon^n\partial_ng_{kl}-\partial_k\varepsilon^ng_{nl}-
\partial_l\varepsilon^ng_{nk}=-{\cal L}_{\varepsilon}g_{kl},
\\
\label{s12}
\delta A_k&=-\varepsilon^m\partial_m A_k-\partial_k\varepsilon^m A_m-\partial_k\varepsilon=
-{\cal L}_{\varepsilon}A_k-\partial_k\varepsilon, 
\\
\label{s02}
\delta \phi &= -\varepsilon^m\partial_m \phi=
-{\cal L}_{\varepsilon}\phi.
\end{align}
Eqs. (\ref{s22})-(\ref{s02})
reproduce standard lower spin gauge transformations.

The length of the geodesic is given by (\ref{els})
\begin{equation}
\label{lsl}
s=\int ds=\int edt=\int{\sqrt{\frac{-g_{mn}\dot x^m\dot x^n}{-2\phi}}dt},
\end{equation}
while the constraint (\ref{constr2}) reads as
\begin{equation}
\label{c2ls}
g_{mn}u^mu^n=2\phi.
\end{equation}
In the case when scalar field takes the background value $-1/2$ (\ref{bgs02})
length (\ref{lsl}) and the constraint (\ref{c2ls}) reproduce standard formulas
\begin{equation*}
ds=\sqrt{-g_{mn}\dot x^m \dot x^n}dt \quad \text{and} \quad g_{mn}\frac{dx^m}{ds}\frac{dx^n}{ds} =-1.
\end{equation*}

Let us finally note that $ds=edt$ is not invariant with respect to
(\ref{hyper}), so the length and all further notions constructed 
out of it are not hyper-Weyl invariant.

\section{Volume}
\label{volume}

In the previous Sections we defined geodesics and length measured by the particle in 
higher spin background. 
In the case of Riemannian geometry once geodesics and length are known one can unambiguously
define all the remaining geometric notions 
 such as volume form, 
parallel transport, scalar curvature, tidal forces, characterized by Riemann tensor etc. 
 One of such quantities is  Einstein-Hilbert action, which is  an integral of the scalar curvature over 
the space, weighted with the volume form. It is tempting to mimic this definitions of 
Riemannian geometry in the case 
of higher spin fields to get a higher spin action. However, these rules
cannot be applied literally to the higher spin case. In particular, the action obtained in this way 
is not gauge invariant. 
To illustrate
 this problem let us consider the volume form. It is supposed to define a measure 
assigned to any infinitesimal set $x\in\Omega$ in the $x$-space. The reasonable extension
of this notion to the  phase space could be to assign volumes to infinitesimal 
cylinders of the form $\Omega\otimes P$ with $x \in\Omega$ and $p$ any,
that is $p_i \in (-\infty,+\infty)$ for all $i$.
However, in general, canonical transformations do not act within a class of such cylinders
mapping them to more general sets in the phase space. Thereby a notion of volume
in $x$ space is not compatible with canonical transformations.
Nonetheless, a subgroup of transformations (\ref{lsgtr1}) still acts within such a 
class of cylinders, so with respect to lower-spin symmetries
 a notion of volume perfectly makes sense.
In the following we derive the literal geometrical generalization 
of Einstein-Hilbert action to the full lower spin sector.

Let us consider a geodesic ball ${\cal B}_x(r)$ with the center $x$ and radius $r$.
A natural way to define the volume form $\omega$ is to demand that the volume of a small geodesic ball 
of radius $r$ in the leading order in $r$ is given by the standard
flat space formula, that is
\begin{equation}
\label{vol1}
\int_{{\cal B}_{x_0}(r)}\omega(x)d^dx=\frac{\pi^{\frac{d}{2}}}{\Gamma (\frac{d}{2}+1)}r^d+{\cal O}(r^{d+1}). 
\end{equation}
Obviously, in this approximation 
one can pull $\omega$ out through the integration sign and evaluate it at the point $x_0$. Hence
\begin{equation}
\label{vol2}
\omega(x_0)=\lim_{r\rightarrow 0}\left(\frac{\pi^{\frac{d}{2}}r^d}{\Gamma (\frac{d}{2}+1)}
\frac{1}{\int_{{\cal B}_{x_0}(r)}d^dx}\right).
\end{equation}

Geodesic ball of radius $r$ consists of a points $x$ such that their geodesic distance $s$ from 
the origin $x_0$ is less than $r$, that is
\begin{equation}
\label{gb1}
{\cal B}_{x_0}(r)=\{x^i:\quad  x^i=x^i_0+u^is+{\cal O}(s^2), \quad \text{where} 
 \quad s<r \quad \text{and} \quad u\in {\cal M}_L \}.
\end{equation}
 One can regard (\ref{gb1}) as a map between  $x$ coordinates inside
the geodesic ball and coordinates $(u,s)$, where $u\in {\cal M}_L$ and $s\in [0,r)$.
Instead, it is more convenient to use a map
\begin{equation}
\label{gb2}
{\cal B}_{x_0}(r)=\{x: \quad x^i = x^i_0 + u^is+{\cal O}(s^2), \quad \text{where}
\quad s=r \quad \text{and}\quad u\in{\cal N}_L \}, 
\end{equation}
where ${\cal N}_L$ is the interior of ${\cal M}_L$, that is 
$\partial{\cal N}_L={\cal M}_L$ or, more explicitly
\begin{equation}
\label{nl}
{\cal N}_L=\{(x,u):\quad u^i\frac{\partial L}{\partial u^i}>L\}.
\end{equation}
The map (\ref{gb2}) allows to compute easily
\begin{equation}
\label{gb3}
\int_{{\cal B}_{x_0}(r)}d^dx=\int_{{\cal N}_L}
\text{det}\left(\frac{\partial x^i}{\partial u^j}\right)d^du=r^d
\int_{{\cal N}_L}d^du +{\cal O}(r^{d+1}).
\end{equation}
Plugging this to (\ref{vol2}) we finally obtain
\begin{equation}
\label{vol3}
\omega(x_0)=\frac{\pi^{\frac{d}{2}}}{\Gamma (\frac{d}{2}+1)}
\frac{1}{\int_{{\cal N}_L}d^du}.
\end{equation}

Analogously, in Hamiltonian approach one can parametrize points of the geodesic ball
by initial momenta
\begin{equation}
\label{gb4}
 x^i=x^i_0+\frac{\partial H}{\partial p_i}s+{\cal O}(s^2).
\end{equation}
As before, we fix $s$ to be $r$ and let $p$ run over the interior
${\cal N}_H$ of ${\cal M}_H$
\begin{equation*}
{\cal N}_H=\{(p,x):\quad H(p,x)<0\}.
\end{equation*}
 Eventually,
we find
\begin{equation}
\label{gb5}
\int_{{\cal B}_{x_0}(r)}d^dx=\int_{{\cal N}_H}
\text{det}\left(\frac{\partial x^i}{\partial p_j}\right)d^dp=r^d
\int_{{\cal N}_H}\text{det}\left(\frac{\partial^2 H}{\partial p_i\partial p_j}\right)d^dp
+{\cal O}(r^{d+1})
\end{equation}
and the volume form is 
\begin{equation}
\label{vol4}
\omega(x_0)=\frac{\pi^{\frac{d}{2}}}{\Gamma (\frac{d}{2}+1)}
\frac{1}{\int_{{\cal N}_H}\text{det}\left(\frac{\partial^2 H}{\partial p_i\partial p_j}\right)d^dp}.
\end{equation}

Let us note that the construction discussed in this Section
 is well-defined only if 
${\cal N}$ is compact. For physically reasonable fields it is not so. For example, 
for Minkowski space ${\cal M}$ is a hyperboloid and ${\cal N}$ is enclosed between
${\cal N}$ and coordinate hypersurfaces. If we will perform literally 
the  computation discussed above we will encounter infinities. Instead of doing so
one can compute as if ${\cal N}$ is compact
 and then analytically continue the result for any values of fields. 
Obviously, it will not spoil transformation properties, that are of the most importance
at this stage. In what follows we will act as if ${\cal N}$ is compact. 
For example, one can consider the action (\ref{lowslag}) with a positive definite $g_{mn}$,
positive $\phi$ and any $A_m$. Then constraint (\ref{c2ls}) defines a compact surface
${\cal M}_L$. Instead of (\ref{lsl}) we have
\begin{equation}
\label{lslpositive}
s=\int{\sqrt{\frac{g_{mn}\dot x^m\dot x^n}{2\phi}}dt}.
\end{equation}
For $\phi=1/2$ one recovers standard Riemannian geometry.

From (\ref{vol3}) one can easily derive that in the lower spin case
\begin{equation}
\label{volls}
\omega(x)=\sqrt{\frac{g}{(2\phi)^d}}.
\end{equation}

Volume form defined in this way transforms as a density with respect to transformations
of the $x$-space. Indeed, the right hand side of (\ref{vol1}) is constant, so
$\omega(x)$ should compensate transformations of $d^dx$, which means that 
$\omega(x)$ is a density. 

As it was already mentioned, transformation properties of $d^dx$
under symplectomorphisms are ill-defined, since $d^dx$ does not contain any 
information about $p$-coordinates of a point, while under canonical transformation
$(x,p)\rightarrow (x',p')$ generically the transformed coordinate $x'$  
depends both on initial $x$ and $p$. More suitable way to define a volume covariantly 
will be discussed in Section \ref{invariants}.

\section{Curvature}
\label{curvature}

In Riemannian geometry the scalar curvature represents the amount by witch the volume of
a geodesic ball ${\cal B}(r)$ deviates from that of the standard ball in Euclidean space in 
subleading orders in $r$. More precisely the volume of ${\cal B}(r)$ in curved space is given
by
\begin{equation}
\label{cur1}
V({\cal B}_{x_0}(r))=\int_{{\cal B}_{x_0}(r)}\omega(x)d^dx.
\end{equation}
The right hand side of (\ref{cur1}) can be decomposed into power series in $r$ and, 
according to the definition of the volume form (\ref{vol1}) the leading term reproduces
the flat space formula. In Riemannian geometry the subleading term vanishes, so the 
first non-trivial term is of order $r^{d+2}$. The scalar curvature ${\cal R}$
is defined as
\begin{equation}
\label{cur2}
V({\cal B}_{x_0}(r))=
\frac{\pi^{\frac{d}{2}}}{\Gamma (\frac{d}{2}+1)}r^d\left(1
-\frac{{\cal R}(x_0)}{6(d+2)}r^2+{\cal O}(r^3)\right).
\end{equation}
We apply this definition to the case of lower spin 
fields and thereby obtain the corresponding generalization of Einstein-Hilbert action.

The computation goes similarly to that of the previous Section. The only difference 
is that now we should compute the integral in the right hand side of (\ref{cur1}) in the leading three
orders in $r$. As before, we perform a change of integration variables from $x^i$ 
to $u^i$ as in (\ref{gb2}) but now we keep further orders in $r$
\begin{equation}
\label{0chcoor1}
x^i=x_0^i+u^ir+v^i(x_0,u)\frac{r^2}{2}+w^i(x_0,u)\frac{r^3}{6}+{\cal O}(r^4),
\end{equation}
where 
\begin{equation*}
v^i = \frac{d^2x^i}{ds^2} \qquad \text{and} \qquad  w^i=\frac{d^3 x^i}{ds^3},
\end{equation*}
which on-shell can be expressed in terms of the initial data 
given by $x_0$ and $u$ by the equation of motion (\ref{eqlag1}). For small enough $r$
 (\ref{0chcoor1}) provides a one-to-one map from $u$ that belong to ${\cal N}_L$  
to the interior of  the geodesic ball $x\in{\cal B}_{x_0}(r)$.

In terms of new integration variables one has
\begin{equation}
\label{curt1}
\int_{{\cal B}_{x_0}(r)}\omega(x)d^dx=\int_{{\cal N}_L}
\omega(x(u))\text{det}\left(\frac{\partial x^i}{\partial u^j}\right)d^du.
\end{equation}
Both determinant and $\omega(x)$ should be decomposed into power series in $r$ 
using (\ref{0chcoor1}). Keeping three leading terms and performing 
integration, one can identify the analog of the curvature.

This computation is performed in details in Appendix for the lower spin case and
 gives (\ref{cur6})  
\begin{equation}
\label{cur7}
{\cal R}={2\phi R}+
(d+2)g^{il}g^{kj}F_{k,l}F_{j,i}+
{2(d-1)}g^{ij}D_iD_j\phi,
\end{equation}
where
\begin{equation*}
F_{m,n}=\frac{1}{2}(\partial_mA_n-\partial_nA_m)
\end{equation*}
and $D_i$ denotes standard covariant derivative.
Then the lower spin sector counterpart of  Einstein-Hilbert action is
\begin{equation}
\label{lsaction}
S[g_{ij},A_i,\phi]=\int{d^dx \sqrt{\frac{g}{(2\phi)^d}}
\left({2\phi R}+
(d+2)g^{il}g^{kj}F_{k,l}F_{j,i}+
{2(d-1)}g^{ij}D_iD_j\phi\right)}.
\end{equation}
In the case of lower spin symmetries $\omega$ transforms as a density, which entails that
$V({\cal B}_{x_0}(r))$ is a scalar. This in turn implies that the curvature defined as
in (\ref{cur2}) is a scalar by construction. For a general canonical transformation the
above reasoning does not work because the volume form itself is ill-defined.

Let us finally note that Minkowski space (\ref{bgs02}) solves equations of motion
derived from (\ref{lsaction}).

\section{Symplectic invariants}
\label{invariants}

Up to now, in order to find the higher spin counterpart of
Einstein-Hilbert action, we tried to mimic the rules of Riemannian geometry.
However, notions of local volume and scalar curvature  proved to be incompatible with symplectic symmetries of
the phase space. In this section we tackle the problem from
 the opposite side: we look at symplectic invariants and choose
  suitable ones. Before doing that, let us review some
  well-known results
 on symplectic invariants (see, for example \cite{hoffzend}).
 
Usually, the term "symplectic invariant" means some measure assigned
to a set ${\cal N}$ in the symplectic space that remains invariant under
symplectic diffeomorphisms. For instance, one can consider the 
phase space
volume 
\begin{equation*}
\int_{\cal N}d^dpd^dx.
\end{equation*} 
 However, it is not the only symplectic invariant.
 
 Whenever one has a compact convex set ${\cal N}$ one can construct
 a Hamiltonian function $H$ such that it vanishes on the boundary
 ${\cal M}$ of ${\cal N}$:
 \begin{equation}
 \label{inv1}
 H(p,x)=0, \quad \forall (p,x)\in {\cal M}.
 \end{equation}
Obviously, such $H$ is not unique. Each of them induces a Hamiltonian
flow on the hypersurface ${\cal M}$. It is easy to see that different
Hamiltonians satisfying (\ref{inv1}) have geodesics that differ only by 
reparametrization. If a Hamiltonian system given by $H$ is integrable, then ${\cal M}$ 
contains $d$ closed geodesics.
 One can profit from reparametrization
freedom so as to make the periods of each closed geodesic equal $2\pi$.
Then, on-shell action on these geodesics provides $d$ 
independent symplectic
invariants\footnote{The invariants that we will construct later,
thereby, are defined only for integrable Hamiltonian systems. Our aim for the
future research is to find an integral form of these invariants, that is to
represent them in a more conventional way as $\int \hat L(H)d^dxd^dp$, where 
$\hat L$ is an operator. It will
allow to define them for any $H$, not necessarily integrable.
The example of such an invariant is a phase-space volume of the set ${\cal N}$,
 which can be expressed in terms of
actions on closed geodesics, but in the same time it can be formulated as
$\int \theta (-H) d^dxd^dp $, which exists independently of integrability of $H$.}.

Let us, however, note that the problem that we are to solve is slightly
different. The  above discussed invariants  depend only on points where
$H=0$. Nothing prevents us from considering more general
invariants, that contain more information about $H$. The reason why
we should not restrict ourselves to the invariants
described above is that such invariants have  extra 
hyper-Weyl symmetry, which is inherent in 
symplectic higher spin fields only \cite{Segal:2000ke,Segal:2002gd}.

To find  suitable invariants let us consider a case of quadratic
Hamiltonian with an extra constant piece
\begin{equation}
\label{inv2}
H=\frac{1}{2} d^{ij}p_ip_j+ e^i{}_jp_ix^j+\frac{1}{2}
 f_{ij}x^ix^j-\frac{C}{2}.
\end{equation}
Suppose, the quadratic part of $H$ is positive definite. Then, for positive
$C$ Hamiltonian $H$ defines a non-empty compact set ${\cal N}$. It 
is well-known, that by linear symplectic transformation (\ref{inv2})
can be put into the normal form
\begin{equation}
\label{inv3}
H=\sum_{i=1}^d\omega_j\frac{p^2_j+x^2_j}{2}-\frac{C}{2}.
\end{equation}
The associated system is the harmonic oscillator whose
equations of motion read
\begin{equation}
\label{inv4}
\frac{d  x_j}{ds}=\omega_j p_j, \qquad \frac{d  p_j}{ds}=-\omega_j x_j.
\end{equation}
If a frequencies vector is non-resonant, that is if
\begin{equation*}
\sum_{i=1}^nk_i\omega_i\ne 0 \qquad \forall k\in \mathbb Z 
\backslash
\{0\}
\end{equation*}
then (\ref{inv4}) on a constraint surface $H=0$ has only $d$ closed orbits. For each $i$ one
has a circular geodesic in $(p_i,x_i)$-plane with a frequency $\omega_i$,
or, equivalently, of length 
\begin{equation*}
L_i=2\pi/\omega_i
\end{equation*}
 (in the sense 
of Section \ref{length}). Lengths of closed geodesics provide another
sort of symplectic invariants, which in the same time do not
posses hyper-Weyl symmetry.

The volume form (\ref{vol4}) for (\ref{inv3}) is
\begin{equation*}
\omega(x)\sim (\omega_1\omega_2\dots \omega_d)^{-1/2}
({C}-\sum_{i=1}^d\omega_j x^2_j)^{-d/2}.
\end{equation*}
It can be integrated over $\sum_{i=1}^d\omega_j x^2_j<C$
(this is a projection of ${\cal N}$ to the $x$ space)
to give
\begin{align}
\notag
V\sim &\int_{\sum_{i=1}^d\omega_j x^2_j<C}d^dx 
(\omega_1\omega_2\dots \omega_d)^{-1/2}
({C}-\sum_{i=1}^d\omega_j x^2_j)^{-d/2}
\\
\label{inv5}
=&
\frac{1}{\omega_1\dots\omega_d}\int_{\sum_{i=1}^d y^2_j<1}
\frac{d^dy}{(1-\sum_{i=1}^d y^2_i)^{d/2}} 
\end{align}
By dropping the infinite overall factor coming from the integration
we find that
\begin{equation}
\label{inv6}
V\sim \frac{1}{\omega_1\dots\omega_d}\sim
L_1\dots L_d.
\end{equation}
Thereby, the total volume is naturally defined as a product of lengths
of closed geodesics. The definition of the total volume (\ref{inv6})
has an advantage that it is manifestly symplectically invariant.

Let us now  plug the ansatz (\ref{inv3}) to  the action (\ref{lsaction}).
The only term that survives in the brackets is 
$g^{ij}D_iD_j\phi\sim \sum_{i=1}^d \omega^2_i$. It is constant, so
one can pull it out the integration sign. Eventually for the action 
(\ref{lsaction}) one finds
\begin{equation}
\label{inv7}
S\sim \frac{\sum_{i=1}^d \omega^2_i}{\omega_1\dots\omega_d}\sim
L_1\dots L_d\left(\sum_{i=1}^d (1/L_i)^2\right).
\end{equation}

These formulas admit a straightforward extension to integrable Hamiltonian systems.
Indeed, in this case 
it is possible to make a canonical change of coordinates
$(p,x)\rightarrow (I,\theta)$ such
that one has $H=H(I)$ and Hamiltonian equations read
\begin{equation}
\label{inv8}
\frac{dI_j}{ds}=0, \quad \frac{d\theta_j}{ds}=\frac{\partial H}{\partial I_j}.
\end{equation}
Variables $(I,\theta)$ are called action-angle variables. Physically, 
the existence of these coordinates can be treated as the generalized
equivalence principle, which implies that higher spin forces can be
locally gauged away by an appropriate change of coordinates.

For each $i$, $1\le i \le d$ there is a unique closed geodesic in the
plane $(I_i,\theta_i)$. Indeed, 
let us introduce
\begin{equation}
\label{inv9}
h_i(I_i)=H(0,\dots, 0,I_i,0,\dots,0).
\end{equation}
With some monotonicity assumptions the constraint $h_i(I_i)=0$
has a unique solution $I_i=I_i^{0}$. From (\ref{inv8}) one sees that
$I_i$ is conserved, while
\begin{equation}
\label{inv10}
\frac{d\theta_j}{ds}=\left.\frac{d h_i}{d I}\right|_{I_i=I_i^0}\equiv
 \omega_i.
\end{equation}
Recalling that the angle variables are cyclic $\theta_i+2\pi=\theta_i$ 
one finds that $\omega_i$ of (\ref{inv10}) is a frequency of 
motion of a point particle along a closed geodesic in $(I_i,\theta_i)$ plane.
One can use these frequencies for any $i$ to find the total volume
(\ref{inv6}) of the space in general higher spin background 
and the higher spin counterpart of Einstein-Hilbert action (\ref{inv7}).

\section*{Conclusion}

Hints that coordinates and momenta should be treated on the same footing 
are present in physics for a long time. In particular, it is known from
classical
mechanics that  trajectories of particles are uniquely determined by initial
coordinates and velocities or, equivalently, by initial coordinates and 
momenta. Canonical commutation relations of quantum mechanics
 rest on the notion of  phase space where coordinates
and momenta are equal in rights.
These theories have a large symmetry, which 
is a group of canonical transformations in the classical case.
We argue, that any manipulations to be covariant 
with respect to canonical transformations 
naturally require higher spin fields in the same way as 
metric  assures covariance
of Riemannian geometry and Einstein
 gravity with respect to diffeomorphisms.
Our ultimate goal was to extend all the objects of Riemannian geometry
such as geodesic, length, volume, curvature etc. 
to a phase space in a symplectically covariant way.

In order to do that we considered the most general action for a relativistic
point particle that depends on coordinates and velocities. Reparametrization
invariance is achieved in a standard way by coupling to an auxiliary world-line
metric.
By identifying Hamiltonian with the generating function for higher spin
fields this action admits interpretation as an action of a point particle
in the higher spin background. Canonical transformations act as higher spin
symmetries. 

Having a point particle action in our disposal we receive access to 
first geometrical notions we aimed for: geodesics and length, which is
measured by the world-line metric.
Knowing them both in Riemannian geometry one can derive all that remains.
In particular, the volume form should be defined so as the volume of an infinitesimal
geodesic ball  depends in a standard way on its radius. The scalar curvature can
be extracted from subleading orders of this dependence. However, it is clear, that
these rules cannot be mimicked if we are aiming at covariance with respect to 
higher spin symmetry. The reason is that already the notion of a set in $x$
 space is not compatible with
canonical transformations. As a result its volume is also ill-defined.
Out of that one can conclude that some of the notions we used to have
in Riemannian geometry undergo essential changes once we go to 
symplectic geometry. 

Despite listed difficulties for general canonical transformations, the
local volume and the curvature
perfectly make sense for gauge symmetries associated with lower spin transformations.
We compute the volume form, the curvature and the counterpart of Einstein-Hilbert
action for a joint sector of spin zero, one and two fields. We expect that this action
is the same as a truncation of the full action containing higher spin fields
 to the lower spin sector.

We also propose the candidates for total volume of the space and the
higher spin counterpart of Einstein-Hilbert action. They are both
defined in terms of lengths (or, equivalently, frequencies)
of closed geodesics. It would be important to reformulate them
in more conventional terms such as in terms of integrals over
the phase space or, probably, over the surface in it. This would
allow to check if the given  action can be used to describe massless higher spin
fields. Let us mention that the ansatz we used to find the symplectic 
invariant candidate for the action does not really probe the Einstein-Hilbert part
of the lower spin action. So, strictly speaking, the invariant we found 
rather generalizes the action of the scalar field than Einstein-Hilbert one.

\section*{Acknowledgments}

I am grateful to N.~Boulanger, E.~Skvortsov, P.~Sundell, V.~Didenko,
D.~Chialva, K.~Siampos, G.~Lucena Gomez, E.~Joung,
  A.~Artsukevich,
M.~Valenzuela and N.~Colombo
for stimulating discussions. I also thank
M.~Vasiliev for interesting comments at the early stage of the work
 and for pointing out references \cite{Segal:2002gd,Bekaert:2010ky}.
 I am grateful to X. Bekaert for reading the manuscript and giving valuable comments.
 I acknowledge the organizers of the workshop "Higher Spins, Strings and Duality"
 where this work was partially performed.
This research  was supported in part by
an ARC contract No. AUWB-2010-10/15-UMONS-1.

\section*{Appendix}
\label{app}

Here we compute in details the volume of a geodesic ball ${\cal B}_{x_0}(r)$
 in the first three orders in radius $r$ in lower spin background.
  To this end we should compute the integral 
 \begin{equation*}
 V({\cal B}_{x_0}(r))=\int_{{\cal N}_L}
\omega(x(u))\text{det}\left(\frac{\partial x^i}{\partial u^j}\right)d^du,
 \end{equation*}
 where 
 \begin{equation}
\label{chcoor1}
x^i=x_0^i+u^ir+v^i(x_0,u)\frac{r^2}{2}+w^i(x_0,u)\frac{r^3}{6}+{\cal O}(r^4),
\end{equation}
\begin{equation*}
v^i = \frac{d^2x^i}{ds^2} \qquad \text{and} \qquad  w^i=\frac{d^3 x^i}{ds^3}.
\end{equation*}
 In the lower spin case $\omega(x)$ is (\ref{volls})
 \begin{equation*}
\omega(x)=\sqrt{\frac{g}{(2\phi)^d}},
 \end{equation*}
 ${\cal N}_L$ is given by $u$ that satisfy (\ref{nl})
 \begin{equation}
\label{lsconstr}
\frac{1}{2}g_{mn}u^mu^n<\phi,
\end{equation}
while $v$ and $w$ read
\begin{align}
\label{2der}
v^i =&-\Gamma^i{}_{,mn}u^mu^n+2g^{ij}F_{j,m}u^m+g^{ij}\partial_j\phi,
\\
\notag
w^i=&-\partial_{r}\Gamma^{i}{}_{,mn}u^ru^mu^n-2\Gamma^{i}{}_{,mn}
(-\Gamma^m{}_{,kl}u^ku^l+2g^{mt}F_{t,k}u^k+g^{mt}\partial_t\phi)u^n\\
\notag
+&2\partial_n g^{ij}u^nF_{j,m}u^m+2g^{ij}\partial_nF_{j,m}u^nu^m
+2g^{ij}F_{j,m}(-\Gamma^m{}_{,kl}u^ku^l\\
\label{3der}
+&2g^{mt}F_{t,k}u^k+g^{mt}\partial_t\phi)
+
\partial_{m}g^{ij}u^m\partial_j\phi+g^{ij}\partial_m\partial_j\phi u^m,
\end{align}
where
\begin{equation*}
F_{m,n}=\frac{1}{2}(\partial_mA_n-\partial_nA_m), \quad \Gamma^i{}_{,jk}=\frac{1}{2}
g^{im}(\partial_jg_{mk}+\partial_kg_{mj}-\partial_mg_{jk}).
\end{equation*}

It is convenient to single out  the leading order of $V({\cal B}_{x_0}(r))$ in $r$
in the  following way
\begin{equation}
\label{cur3}
V({\cal B}_{x_0}(r))=\int_{{\cal N}_L}
\omega(x(u))\text{det}\left(\frac{\partial x^i}{\partial u^j}\right)d^du=
r^d\int_{{\cal N}_L}
\omega(x(u))Md^du,
\end{equation}
where 
\begin{equation}
\label{M}
M=\text{det}\left(M^i{}_j\right), \qquad
M^i{}_j=\delta^i{}_j+\frac{r}{2}\frac{\partial v^i}{\partial u^j}+
\frac{r^2}{6}\frac{\partial w^i}{\partial u^j}+{\cal O}(r^3).
\end{equation}

Expanding $\omega$ and $M$ into power series in $r$  
\begin{align*}
\omega =& \omega\Big|_{r=0}+\frac{\partial\omega}{\partial x^i}\Big|_{r=0}
(u^ir+\frac{r^2}{2}v^i)+
\frac{1}{2}\frac{\partial^2\omega}{\partial x^i\partial x^j}\Big|_{r=0}u^i ru^j r+{\cal O}(r^3),\\
M=&M\Big|_{r=0}+\frac{dM}{dr}\Big|_{r=0}r+\frac{d^2M}{dr^2}\Big|_{r=0}\frac{r^2}{2}+{\cal O}(r^3),
\end{align*}
and
plugging the result into (\ref{cur3}) in leading orders we obtain
\begin{align}
\notag
V({\cal B}_{x_0}(r))=&
r^d\int_{{\cal N}_L}
(\omega\cdot M)\Big|_{r=0}d^du+r^{d+1}\int_{{\cal N}_L}
\left(\omega\frac{dM}{dr}+\frac{\partial\omega}{\partial x^i}u^iM\right)\Big|_{r=0}d^du
\\
\label{cur5}
+&\frac{r^{d+2}}{2}\int_{{\cal N}_L}
\left(\omega\frac{d^2M}{dr^2}+2\frac{\partial \omega}{\partial x^i}u^i\frac{dM}{dr}+
\frac{\partial^2\omega}{\partial x^i\partial x^j}u^iu^jM+
\frac{\partial \omega}{\partial x^i}v^i M\right)\Big|_{r=0}d^du.
\end{align}
The derivatives of determinants can be written in terms of derivatives
of matrix elements in a standard way
\begin{align*}
\frac{dM}{dr}=&M(M^{-1})^i{}_j\frac{dM^{j}{}_i}{dr}, 
\\
\frac{d^2M}{dr^2}=&M(M^{-1})^i{}_j\frac{dM^{j}{}_i}{dr}(M^{-1})^k{}_l\frac{dM^{l}{}_k}{dr}
\\
-&M(M^{-1})^i{}_j\frac{dM^{j}{}_k}{dr}(M^{-1})^k{}_l\frac{dM^{l}{}_i}{dr}+
M(M^{-1})^i{}_j\frac{d^2M^j{}_i}{dr^2}.
\end{align*}

The integration domain (\ref{lsconstr}) is symmetric with respect to the origin, so
 the odd powers of $u$ drop out under integration. The $u$-independent and quadratic in $u$
terms give
\begin{align}
\label{uv0}
\int_{{\cal N}_L}
d^du&=\int_{\frac{1}{2}g_{mn}u^mu^n<\phi}
d^du=\frac{\pi^{\frac{d}{2}}}{\Gamma (\frac{d}{2}+1)}\sqrt{\frac{(2\phi)^d}{g}}, 
\\
\label{uv2}
\int_{{\cal N}_L}a_{ij}u^iu^j
d^du&=\int_{\frac{1}{2}g_{mn}u^mu^n<\phi}a_{ij}u^iu^j
d^du=\frac{\pi^{\frac{d}{2}}}{\Gamma (\frac{d}{2}+1)}\sqrt{\frac{(2\phi)^d}{g}}
\cdot\frac{2\phi a_{ij}g^{ij}}{d+2}. 
\end{align} 

Now we apply (\ref{uv0}), (\ref{uv2}) and compute  (\ref{cur5}). The  
leading $r^d$ term reproduces the standard flat contribution as enforced by the construction 
of the volume form, the coefficient 
in front of $r^{d+1}$ vanishes, while the remaining integrals read
\begin{align}
\notag
\int_{{\cal N}_L}\left(\omega\frac{d^2M}{dr^2}\right)\Big|_{r=0}d^du=&
\frac{\pi^{\frac{d}{2}}}{\Gamma (\frac{d}{2}+1)}\frac{2\phi}{d+2}
\left(\Gamma^{l}{}_{,lm}\Gamma^j{}_{,jn}+
\frac{1}{3}\Gamma^{i}{}_{,km}\Gamma^k{}_{,in}+
\frac{2}{3}\Gamma^{i}{}_{,ji}\Gamma^j{}_{,mn}
\right.
\\
\notag
-&\left. \frac{1}{3}\partial_i\Gamma^i{}_{,mn}-\frac{2}{3}\partial_m\Gamma^i{}_{,in}
\right) g^{mn}+\frac{\pi^{\frac{d}{2}}}{\Gamma (\frac{d}{2}+1)}
\left(-\frac{1}{4}g^{it}g^{ks}F_{k,t}F_{s,i}\right.
\\
\label{comp1}
-&\frac{2}{3}\Gamma^i{}_{,mi}g^{mt}\partial_t\phi+
\left.\frac{1}{3}\partial_ig^{li}\partial_l\phi+\frac{1}{3}g^{il}\partial_i\partial_l\phi
\right),
\end{align}
\begin{align}
\label{comp2}
\int_{{\cal N}_L}\left(
2\frac{\partial \omega}{\partial x^i}u^i\frac{dM}{dr}\right)\Big|_{r=0}d^du=
\frac{\pi^{\frac{d}{2}}}{\Gamma (\frac{d}{2}+1)}\frac{2\phi}{d+2}
\left(\Gamma^l{}_{,jl}-\frac{d}{2}\frac{\partial_j\phi}{\phi}\right)
\left(-2\Gamma^i{}_{,im}\right)g^{mj},
\end{align}
\begin{align}
\notag
\int_{{\cal N}_L}\left(
\frac{\partial^2\omega}{\partial x^i\partial x^j}u^iu^jM\right)\Big|_{r=0}d^du=&
\frac{\pi^{\frac{d}{2}}}{\Gamma (\frac{d}{2}+1)}\frac{2\phi}{d+2}
\left[\partial_j\Gamma^l{}_{,il}-\frac{d}{2}\left(\frac{\partial_i\partial_j\phi}{\phi}
-\frac{\partial_i\phi\partial_j\phi}{\phi^2}\right)\right.
\\
\label{comp3}
+&\left.\left(\Gamma^l{}_{,il}-\frac{d}{2}\frac{\partial_i\phi}{\phi}\right)
\left(\Gamma^m{}_{,jm}-\frac{d}{2}\frac{\partial_j\phi}{\phi}\right)\right]g^{ij},
\end{align}
\begin{align}
\notag
\int_{{\cal N}_L}\left(
\frac{\partial \omega}{\partial x^i}v^i Md^du\right)\Big|_{r=0}=&
\frac{\pi^{\frac{d}{2}}}{\Gamma (\frac{d}{2}+1)}\frac{2\phi}{d+2}
\left(\Gamma^l{}_{,il}-\frac{d}{2}\frac{\partial_i\phi}{\phi}\right)
\left(-\Gamma^i{}_{,mn}g^{mn}\right)
\\
\label{comp4}
+&\frac{\pi^{\frac{d}{2}}}{\Gamma (\frac{d}{2}+1)}
\left(\Gamma^l{}_{,il}-\frac{d}{2}\frac{\partial_i\phi}{\phi}\right)g^{ij}\partial_i\phi.
\end{align}
Plugging (\ref{comp1})-(\ref{comp4}) into (\ref{cur5}) we finally find
\begin{align}
\notag
V({\cal B}_{x_0}(r))=& \frac{\pi^{\frac{d}{2}}}{\Gamma (\frac{d}{2}+1)}
r^d\left[1-\frac{r^2}{2}\Big(\frac{2\phi R}{3(d+2)}\right.\\
\label{cur6}
+&\left. \frac{1}{3}g^{il}g^{kj}F_{k,l}F_{j,i}+
\frac{2(d-1)}{3(d+2)}g^{ij}D_iD_j\phi\Big)\right]+{\cal O}(r^{d+3}),
\end{align}
where $R$ is the standard scalar curvature of gravitational field and $D_i$ is a covariant derivative.

\providecommand{\href}[2]{#2}\begingroup\raggedright\endgroup


\begin{thebibliography}{10}

\bibitem{Sorokin:2004ie}
D.~Sorokin, ``{Introduction to the classical theory of higher spins},'' {\em
  AIP Conf.Proc.} {\bf 767} (2005) 172--202,
\href{http://arXiv.org/abs/hep-th/0405069}{{\tt hep-th/0405069}}.

\bibitem{Bekaert:2005vh}
X.~Bekaert, S.~Cnockaert, C.~Iazeolla, and M.~Vasiliev, ``{Nonlinear higher
  spin theories in various dimensions},''
\href{http://arXiv.org/abs/hep-th/0503128}{{\tt hep-th/0503128}}.

\bibitem{Bekaert:2010hw}
X.~Bekaert, N.~Boulanger, and P.~Sundell, ``{How higher-spin gravity surpasses
  the spin two barrier: no-go theorems versus yes-go examples},'' {\em
  Rev.Mod.Phys.} {\bf 84} (2012) 987--1009,
\href{http://arXiv.org/abs/1007.0435}{{\tt 1007.0435}}.

\bibitem{Sagnotti:2011qp}
A.~Sagnotti, ``{Notes on Strings and Higher Spins},''
\href{http://arXiv.org/abs/1112.4285}{{\tt 1112.4285}}.

\bibitem{Vasiliev:1990en}
M.~A. Vasiliev, ``{Consistent equation for interacting gauge fields of all
  spins in (3+1)-dimensions},'' {\em Phys.Lett.} {\bf B243} (1990)
378--382.

\bibitem{Vasiliev:2003ev}
M.~Vasiliev, ``{Nonlinear equations for symmetric massless higher spin fields
  in (A)dS(d)},'' {\em Phys.Lett.} {\bf B567} (2003) 139--151,
\href{http://arXiv.org/abs/hep-th/0304049}{{\tt hep-th/0304049}}.

\bibitem{Didenko:2006zd}
V.~Didenko, A.~Matveev, and M.~Vasiliev, ``{BTZ Black Hole as Solution of 3-D
  Higher Spin Gauge Theory},'' {\em Theor.Math.Phys.} {\bf 153} (2007)
  1487--1510,
\href{http://arXiv.org/abs/hep-th/0612161}{{\tt hep-th/0612161}}.

\bibitem{Didenko:2009td}
V.~Didenko and M.~Vasiliev, ``{Static BPS black hole in 4d higher-spin gauge
  theory},'' {\em Phys.Lett.} {\bf B682} (2009) 305--315,
\href{http://arXiv.org/abs/0906.3898}{{\tt 0906.3898}}.

\bibitem{Gutperle:2011kf}
M.~Gutperle and P.~Kraus, ``{Higher Spin Black Holes},'' {\em JHEP} {\bf 1105}
  (2011) 022,
\href{http://arXiv.org/abs/1103.4304}{{\tt 1103.4304}}.

\bibitem{Iazeolla:2011cb}
C.~Iazeolla and P.~Sundell, ``{Families of exact solutions to Vasiliev's 4D
  equations with spherical, cylindrical and biaxial symmetry},'' {\em JHEP}
  {\bf 1112} (2011) 084,
\href{http://arXiv.org/abs/1107.1217}{{\tt 1107.1217}}.

\bibitem{Castro:2011iw}
A.~Castro, R.~Gopakumar, M.~Gutperle, and J.~Raeymaekers, ``{Conical Defects in
  Higher Spin Theories},'' {\em JHEP} {\bf 1202} (2012) 096,
\href{http://arXiv.org/abs/1111.3381}{{\tt 1111.3381}}.

\bibitem{Iazeolla:2012nf}
C.~Iazeolla and P.~Sundell, ``{Biaxially symmetric solutions to 4D higher-spin
  gravity},''
\href{http://arXiv.org/abs/1208.4077}{{\tt 1208.4077}}.

\bibitem{Ammon:2012wc}
M.~Ammon, M.~Gutperle, P.~Kraus, and E.~Perlmutter, ``{Black holes in three
  dimensional higher spin gravity: A review},''
\href{http://arXiv.org/abs/1208.5182}{{\tt 1208.5182}}.

\bibitem{Berends:1984rq}
F.~A. Berends, G.~Burgers, and H.~van Dam, ``{ON THE THEORETICAL PROBLEMS IN
  CONSTRUCTING INTERACTIONS INVOLVING HIGHER SPIN MASSLESS PARTICLES},'' {\em
  Nucl.Phys.} {\bf B260} (1985)
295.

\bibitem{Segal:2000ke}
A.~Y. Segal, ``{Point particle in general background fields and generalized
  equivalence principle},''
\href{http://arXiv.org/abs/hep-th/0008105}{{\tt hep-th/0008105}}.

\bibitem{Segal:2002gd}
A.~Y. Segal, ``{Conformal higher spin theory},'' {\em Nucl.Phys.} {\bf B664}
  (2003) 59--130,
\href{http://arXiv.org/abs/hep-th/0207212}{{\tt hep-th/0207212}}.

\bibitem{Vasiliev:2005zu}
M.~Vasiliev, ``{Actions, charges and off-shell fields in the unfolded dynamics
  approach},'' {\em Int.J.Geom.Meth.Mod.Phys.} {\bf 3} (2006) 37--80,
\href{http://arXiv.org/abs/hep-th/0504090}{{\tt hep-th/0504090}}.

\bibitem{Engquist:2005yt}
J.~Engquist and P.~Sundell, ``{Brane partons and singleton strings},'' {\em
  Nucl.Phys.} {\bf B752} (2006) 206--279,
\href{http://arXiv.org/abs/hep-th/0508124}{{\tt hep-th/0508124}}.

\bibitem{Grigoriev:2006tt}
M.~Grigoriev, ``{Off-shell gauge fields from BRST quantization},''
\href{http://arXiv.org/abs/hep-th/0605089}{{\tt hep-th/0605089}}.

\bibitem{Bekaert:2009ud}
X.~Bekaert, E.~Joung, and J.~Mourad, ``{On higher spin interactions with
  matter},'' {\em JHEP} {\bf 0905} (2009) 126,
\href{http://arXiv.org/abs/0903.3338}{{\tt 0903.3338}}.

\bibitem{Bekaert:2010ky}
X.~Bekaert, E.~Joung, and J.~Mourad, ``{Effective action in a higher-spin
  background},'' {\em JHEP} {\bf 1102} (2011) 048,
\href{http://arXiv.org/abs/1012.2103}{{\tt 1012.2103}}.

\bibitem{NeveuLPI}
A.~Neveu, ``A stepping stone between einstein-yang-mills and strings?.'' a talk
  given at Lebedev Physical Institute, May, 2011.

\bibitem{Grigoriev:2012xg}
M.~Grigoriev, ``{Parent formulations, frame-like Lagrangians, and generalized
  auxiliary fields},'' {\em JHEP} {\bf 1212} (2012) 048,
\href{http://arXiv.org/abs/1204.1793}{{\tt 1204.1793}}.

\bibitem{Bars:1998gv}
I.~Bars and C.~Deliduman, ``{Gauge symmetry in phase space with spin: A Basis
  for conformal symmetry and duality among many interactions},'' {\em
  Phys.Rev.} {\bf D58} (1998) 106004,
\href{http://arXiv.org/abs/hep-th/9806085}{{\tt hep-th/9806085}}.

\bibitem{deWit:1979pe}
B.~de~Wit and D.~Z. Freedman, ``{Systematics of Higher Spin Gauge Fields},''
  {\em Phys.Rev.} {\bf D21} (1980)
358.

\bibitem{Fronsdal:1978rb}
C.~Fronsdal, ``{Massless Fields with Integer Spin},'' {\em Phys.Rev.} {\bf D18}
  (1978)
3624.

\bibitem{Vasiliev:1980as}
M.~A. Vasiliev, ``{'Gauge' Form of Description of Massless Fields with Arbitrary
  Spin},'' {\em Sov.J.Nucl.Phys.} {\bf 32} (1980) 439 [{\em Yad.Fiz.} {\bf 32} (1980) 855].

\bibitem{Pashnev:1998ti}
A.~Pashnev and M.~Tsulaia, ``{Description of the higher massless irreducible
  integer spins in the BRST approach},'' {\em Mod.Phys.Lett.} {\bf A13} (1998)
  1853--1864,
\href{http://arXiv.org/abs/hep-th/9803207}{{\tt hep-th/9803207}}.

\bibitem{Buchbinder:2001bs}
I.~Buchbinder, A.~Pashnev, and M.~Tsulaia, ``{Lagrangian formulation of the
  massless higher integer spin fields in the AdS background},'' {\em
  Phys.Lett.} {\bf B523} (2001) 338--346,
\href{http://arXiv.org/abs/hep-th/0109067}{{\tt hep-th/0109067}}.

\bibitem{Francia:2002aa}
D.~Francia and A.~Sagnotti, ``{Free geometric equations for higher spins},''
  {\em Phys.Lett.} {\bf B543} (2002) 303--310,
\href{http://arXiv.org/abs/hep-th/0207002}{{\tt hep-th/0207002}}.

\bibitem{Bekaert:2002dt}
X.~Bekaert and N.~Boulanger, ``{Tensor gauge fields in arbitrary
  representations of GL(D,R): Duality and Poincare lemma},'' {\em
  Commun.Math.Phys.} {\bf 245} (2004) 27--67,
\href{http://arXiv.org/abs/hep-th/0208058}{{\tt hep-th/0208058}}.

\bibitem{deMedeiros:2002ge}
P.~de~Medeiros and C.~Hull, ``{Exotic tensor gauge theory and duality},'' {\em
  Commun.Math.Phys.} {\bf 235} (2003) 255--273,
\href{http://arXiv.org/abs/hep-th/0208155}{{\tt hep-th/0208155}}.

\bibitem{Francia:2002pt}
D.~Francia and A.~Sagnotti, ``{On the geometry of higher spin gauge fields},''
  {\em Class.Quant.Grav.} {\bf 20} (2003) S473--S486,
\href{http://arXiv.org/abs/hep-th/0212185}{{\tt hep-th/0212185}}.

\bibitem{Bekaert:2003az}
X.~Bekaert and N.~Boulanger, ``{On geometric equations and duality for free
  higher spins},'' {\em Phys.Lett.} {\bf B561} (2003) 183--190,
\href{http://arXiv.org/abs/hep-th/0301243}{{\tt hep-th/0301243}}.

\bibitem{Milgrom:1983ca}
M.~Milgrom, ``{A Modification of the Newtonian dynamics as a possible
  alternative to the hidden mass hypothesis},'' {\em Astrophys.J.} {\bf 270}
  (1983)
365--370.

\bibitem{hoffzend}
H.~Hofer and E.~Zender, {\em Symplectic invariants and Hamiltonian dynamics}.
\newblock Modern Birkhauser Classics, 1994.

\end{thebibliography}
\end{document}